\documentclass{kapproc} % Computer Modern font calls
\input psfig.tex

\let\footnote\savefootnote

\kluwerbib
\let\lcitebracket(
\let\rcitebracket)

\newcommand{\bi}[1]{\mbox{\boldmath ${#1}$}}

\newcommand{\lesssim}{\mbox{\hskip 2pt \raisebox{2.7pt}{$<$}\hskip -8pt
\raisebox{-2.6pt}{$\sim$}\hskip 2pt}}

\newcommand{\pll}{\parallel}

\newcommand{\rmi}{{\rm i}}

\newcommand{\Ree}{\mathop{{\Re}{\rm e}}}
\newcommand{\Imm}{\mathop{{\Im}{\rm m}}}
\newcommand{\half}{{\textstyle{\frac{1}{2}}}}
\newcommand{\quarter}{{\textstyle{\frac{1}{4}}}}

\newcommand{\up}{\uparrow}
\newcommand{\dn}{\downarrow}

\newcommand{\pvint}{\mathop{\cal P}\!\!\!\int}

\newcommand{\stbrkt}[1]{\left|{#1}\right|}

\newcommand{\bra}[1]{\left<#1\right|}
\newcommand{\ket}[1]{\left|#1\right>}
\newcommand{\braket}[2]{\left<\left.#1\right|#2\right>}

\newcommand{\boxit}[1]{\vbox{\hrule height1pt \hbox{\vrule 
width1pt\kern3pt\vbox
{\kern3pt#1\kern3pt}\kern3pt\vrule width1pt}\hrule height1pt}}
\newcommand{\invisibleboxit}[1]{\vbox{ height1pt \hbox{ width1pt\kern3pt\vbox
{\kern3pt#1\kern3pt}\kern3pt\vrule width1pt} height1pt}}

\begin{document}

\articletitle{Berry Phase with Environment:\\ Classical versus Quantum}

\articlesubtitle{}

\author{Robert S. Whitney}
\affil{D\'epartement de Physique Th\'eorique, Universit\'e de Gen\`eve,\\
        CH-1211 Gen\`eve 4, Switzerland}
\email{}

\author{Yuriy Makhlin}
\affil{\hbox{Institut  f\"ur  Theoretische Festk\"orperphysik,
Universit\"at   Karlsruhe,  76128   Karlsruhe, Germany}\\
Landau Institute for Theoretical Physics, Kosygin st. 2,
117940 Moscow, Russia}
\email{}

\author{Alexander Shnirman}
\affil{Institut  f\"ur  Theoretische Festk\"orperphysik\\
Universit\"at   Karlsruhe,  76128   Karlsruhe, Germany}
\email{}

\author{Yuval Gefen}
\affil{Department of  Condensed Matter  Physics\\
The Weizmann  Institute of Science,  Rehovot  76100,  Israel}
\email{}

\begin{abstract}
We discuss the concept of the Berry phase in a dissipative system. 
We show that 
one can identify a Berry phase in a weakly-dissipative system and find the 
respective correction to this quantity, induced by the environment. This 
correction is expressed in terms of the symmetrized noise power and is 
therefore insensitive to the nature of the noise representing the environment,
namely whether it is classical or quantum mechanical.
It is only the  spectrum of the noise which counts.  
We analyze a model of a spin-half (qubit) anisotropically coupled to 
its environment and explicitly show the coincidence between 
the effect of a quantum environment and a classical one.
\end{abstract}


%%%%%%%%%%%%%%%%%%%%%%%%%%%%%%%%%%%%%%%%%%%%%%%%%%%%%%%%%%%%%%%%
%%%%%%%%%%%%%%%%%%%%%%%%%%%%%%%%%%%%%%%%%%%%%%%%%%%%%%%%%%%%%%%%
\section*{Introduction}

Three papers published independently in 1932 by Zener, Landau and
Stueckelberg \cite{LZ-Landau,LZ-Zener,LZ-Stueckelberg} have introduced the 
phenomenon known
today as Landau-Zener  tunneling. The idea is  to consider a
2-level system, where the energy of each level varies  linearly
with a classical variable (which, in turn, is varied linearly in
time). As function of time, $t$, the energy levels should
intersect but for the  inter-level coupling $\Delta$ which  gives
rise to an ``avoided crossing'' in the spectrum, cf. Fig.1. Using the spin 
notation, one can write the Hamiltonian as $\hat{\cal
H} = \alpha t S_z + \Delta S_x$. Here $\bi{S}=\bi{\sigma}/2$, and
$\sigma_z,\sigma_x$ are Pauli spin-$1/2$ operators; $\alpha$ is the rate of 
change of the energy of the
pseudo-spin at asymptotic times. The avoided crossing gap is
$\Delta$. The probability of transition from, say, the lower level at time 
$-\infty$ , to the upper level at time $+\infty$ is given by
$P_{\rm LZ}=\exp[-(\pi/2)\Delta^2/\alpha]$.

Besides being ubiquitous in physics and chemistry, the Landau-Zener
framework appears to suggest a natural definition for the
notion of adiabaticity. The adiabatic  limit is approached when  $P_{\rm 
LZ}<<1$, i.e., $\alpha<<\Delta^2$. The latter inequality involves a comparison 
of the rate of change  (of the time dependent term in the Hamiltonian)
with the gap  in the spectrum, $\Delta$. This notion of the adiabatic
limit has become widespread. A closer look suggests that, in
general, adiabaticity \hbox{\bf cannot} be associated with comparing the rate 
of
change to the gap. Indeed, on one hand any finite, discrete-spectrum system
is coupled, however weakly, to the rest of the universe. Hence
the emerging spectrum is, at least in principle, always continuous and
gapless. 
The naive view would then imply that the adiabatic
limit cannot be approached. This, on the other hand cannot  be
correct: if we consider  a finite system with a discrete
spectrum, for which adiabaticity is well defined, it is inconceivable that
an infinitesimal coupling to the continuum (rendering the overall spectrum 
continuous) 
will change its physics in a
dramatic way. The resolution of this problem is provided by the observation 
that 
the criterion for adiabaticity involves not only  spectral properties but 
also  the {\bf matrix elements} of the  system-environment coupling.

To gain some insight into this problem we focus here on the analysis of the
Berry phase \cite{Berry84} in a weakly dissipative system.  It is particularly 
timely to
address this issue now given the recent experimental activities in realization
of controlled quantum two-level systems (qubits), and in particular, the 
interest in
observing a Berry phase (BP) (see, e.g., \cite{Fazio_Berry}).  For instance, the
superconducting qubits have a coupling to their environment, which is weak but
not negligible~\cite{Nakamura99,Vion02,Chiorescu03}, and thus it is important to
find both the conditions under which the Berry phase can be observed and the
nature of that Berry phase.

In this paper we appeal to a simple analysis
of the problem.  We first, in Section~\ref{sec:quantum}, consider a 
quantum-mechanical framework, where a  perturbative approach is taken.  
When the environment is
replaced by a single oscillator, a second-order perturbation analysis is
straightforward and produces a result which allows for  a simple 
interpretation.  We
then generalize the calculation for a host of  environmental modes.  In
Section~\ref{sec:classical} we consider a toy model where the environment is
replaced by a {\bf classical} stochastic force.  The quantities of interest,
the Lamb shift and the Berry phase, are then calculated, and simple heuristic
arguments are given to interpret the results.  
To complete the analogy with the analysis of the previous
section, here the ``single-oscillator environment'' is replaced by a simple
periodic classical force (of random amplitude).  In Section~\ref{sec:concl} we
summarize  the relation between the quantum mechanical approach and the
classical model  in more  general terms.

%%%%%%%%%%%%%%%%%%%%%%%%%%%%%%%%%%%%%%%%%
\section{The system: spin + environment}

We begin in the conventional way by writing the Hamiltonian for
the ``universe'' (system $+$ environment) as
\begin{eqnarray}
\hat{\cal H}= \hat{\cal H}_{\rm syst} +\hat{\cal H}_{\rm env} +
\hat{\cal V}_{\rm coupling}
\end{eqnarray}
The system is defined  as the set of those  quantum degrees of freedom that
one is interested to  control and measure;   the environment consists of all
the rest, namely those degrees-of-freedom   we can neither control nor measure. 
The coupling between the
system and environment is ${\cal V}_{\rm coupling}$.
The properties of the environment are controlled by macroscopic parameters, 
such 
as temperature.
Our treatment below applies to a reservoir at either  zero or a  
finite temperature.

For our purposes it is sufficient to represent  the
environment by a single operator $X$ which  couples to a spin. 
The Hamiltonian then becomes
\begin{equation}
\label{Eq:Ham}
\hat{\cal H}= -\half\, \mu g\, \bi{B}\cdot \hat{\bi{\sigma}}
-\half X \sigma_z + \hat{\cal H}_{\rm env}
\,.
\end{equation}
Hereafter we put $\mu g =1$.
Below we express our results in terms of the statistical properties 
(correlators) of the environment's noise, $X(t)$. 
Depending on the physical situation at hand, one can choose to
model the environment via a bath of harmonic oscillators
\cite{Vernon63,Caldeira83}. In  this case the generalized coordinate  of the 
reservoir
is defined as  $X=\sum \lambda_i x_i$, 
where $\{x_i\}$ are the coordinate operators of the 
oscillators and $\{\lambda_i\}$ are the respective couplings.  
Eq.~\ref{Eq:Ham} is then referred 
to as the spin-boson Hamiltonian ~\cite{Leggett87}.
Another example of a reservoir could be 
a spin bath~\cite{Spin_Bath}~\footnote{For any 
reservoir in equilibrium the fluctuation-dissipation 
theorem provides the relation between the symmetrized and antisymmetrized 
correlators of the noise: $S_X(\omega) = A_X(\omega)\coth(\omega/2T)$.
Yet, the temperature dependence of $S_X$ and $A_X$ may vary depending on the 
type of the environment. For an oscillator bath, $A_X$ (also called the spectral 
density $J_X(\omega)$) is temperature-independent, so that $S_X(\omega) = 
J_X(\omega)\coth(\omega/2T)$. On the other hand, for a 
spin bath $S_X$ is temperature-independent and is related 
to the spins' density of states, while $A_X(\omega) = 
S_X(\omega)\tanh(\omega/2T)$.}.
%%%%%%%%%%%%%%%%%%%%%%%%%%%
However, in our analysis below  we do not specify the type  of the environment. 
We will only assume that the reservoir gives rise to markovian evolution on the 
time scales of interest. More specifically, the evolution is markovian at time 
scales longer than a certain characteristic time $\tau_{\rm c}$, determined by 
the environment~\footnote{This time may be given by the correlation time of the 
fluctuations, but in general is a more subtle characteristic of the spectrum
related to its roughness near qubit's frequencies. Note further that for 
singular spectra $\tau_{\rm c}$ may be ill defined and the perturbative analysis 
may fail. See, e.g., \cite{Bloch_Derivation,Redfield_Derivation,%
Slichter,Our_Erice,FrankW-Osc,wg-longpaper}.}.
%%%%%%%%%%%%%%%%%%%%%%%%
We  assume that $\tau_{\rm c}$ is shorter than the  dissipative time scales 
introduced by the environment, such as the dephasing or relaxation times and the 
inverse Lamb shift (the scale of the shortest of which 
we denote as $T_{\rm diss}$, $\tau_{\rm c}\ll T_{\rm diss}$). We further assume 
that $\tau_{\rm c}\ll t_{\rm P}$, the characteristic variation time of the field 
${\bf B}(t)$.
Moreover, under these conditions we may
consider only lowest-order (in the system-environment coupling) 
contributions  to the quantities of interest: energy shifts, BP and
relaxation rates. Indeed, if one divides the 
evolution time interval into short domains ($\ll t_{\rm P}$), longer than 
$\tau_{\rm c}$ but 
shorter than $T_{\rm diss}$, fluctuations at different domains are uncorrelated 
and their effect can be analyzed separately. At the same time, for each domain 
($\ll T_{\rm diss}$) the effect of the noise is 
weak. Thus, to the leading order corrections to the 
dynamics may be described as corrections to the rates (energies) of the spin 
dynamics, which may be estimated perturbatively.
We also consider an underdamped spin, with the dissipative times longer than the 
period of the coherent dynamics, $T_{\rm diss}\gg 1/B$. 
This implies that the time windows alluded to above consist of numerous 
oscillations, in other words they are $\gg 1/B$.

We have chosen an anisotropic spin-environment coupling, $\propto\sigma_z$.
This is a realistic model, e.g., for many designs of solid-state qubits, where 
the different components of  the ``spin''   are influenced by entirely 
different environmental degrees of 
freedom \cite{Nakamura99,Vion02,Chiorescu03}. While our analysis can be 
generalized to account for multiple-directional fluctuating fields
\cite{wmsg-geometricBP}, 
here we focus  on 
unidirectional fluctuations (along  the $z$ axis).

Another remark to be made concerns the possibility to observe a (weak)
dissipative correction to Berry phase in spite of the dephasing and relaxation
phenomena.  While the respective time scales ($T_1$, $T_2$ and the inverse of
the correction to the Berry phase) scale similarly with the strength of
fluctuations (inversely proportionally to the noise power), they are dominated
by different frequency domains.  Indeed, the dephasing and relaxation are known
to be dominated by resonant fluctuations with frequencies close to $B$ (for the
relaxation and the corresponding contribution to dephasing) and $0$ (for the
pure dephasing), cf.~Eq.~(\ref{Eq:T2}) below.  In contrast, as we shall see
below, the Lamb shift and the correction to the Berry phase accumulate
contribution from the entire range of frequencies.  Thus, one may think of
(engineering) a system with an environment whose fluctuations at $\nu\sim B$
and $\nu\sim 0$ are suppressed.  In this case, one can easily observe an
observable correction to the Berry phase at times when the dephasing and
relaxation are still negligible.

%%%%%%%%%%%%%%%%%%%%%%%%%%%%%%%%%%%%%%%%%%%%%%%%%%%%%%%%%%%%%%%%%%%%%
%%%%%%%%%%%%%%%%%%%%%%%%%%%%%%%%%%%%%%%%%%%%%%%%%%%%%%%%%%%%%%%%%%%%%
\section{Quantum-mechanical analysis}
\label{sec:quantum}

In this section we consider a two-level system coupled to an environment which 
we treat as a quantum-mechanical system. We begin with a discussion of the Lamb 
shift and then show, in Subsection~\ref{sec:LStoBP}, how the results for the 
Lamb shift may be used to find the environment-induced correction to the Berry 
phase and the relaxation times.

%%%%%%%%%%%%%%%%%%%%%%%%%%%%%%%%%%%%%%%%%%%%%%%%%%%%%%%%%%%%%%%%%%%%%
\subsection{Lamb shift as level repulsion}

Consider first, for illustration, a simple system of the spin coupled to a 
single oscillator, with the Hamiltonian
\begin{equation}
{\cal H}= -\half B\sigma_z -\half c \sigma_x (a^{\dag} + a) + 
\omega_0 a^{\dag}a
\ ,
\end{equation}
where $c$ is the coupling constant. Let $\ket{n}$ denote the $n$-th level of 
the 
oscillator; the second-order corrections to the energies of the states 
$\ket{\up,0}$ and $\ket{\dn,0}$ are
\begin{equation}
E_{\up}^{(2)} = - \frac{\stbrkt{\bra{\up,0}{\cal V}\ket{\dn,1}}^2}%
{\omega_0+B} = -\frac{1}{4}\,\frac{c^2}{\omega_0+B}
\,,
\label{Eq:dEup} 
\end{equation} 
and
\begin{equation}
E_{\dn}^{(2)} = - \frac{\stbrkt{\bra{\dn,0}{\cal V}\ket{\up,1}}^2}%
{\omega_0-B} = -\frac{1}{4}\,\frac{c^2}{\omega_0-B}
\,, 
\label{Eq:dEdn} 
\end{equation} 
where ${\cal V} \equiv (c/2) \,\sigma_x\,(a^{\dag} + a)$ is the 
perturbation. This results in the following correction to the level spacing 
$E_{\dn} - E_{\up}$:
\begin{equation}
E_{\dn}^{(2)} - E_{\up}^{(2)} = 
\frac{c^2}{2}\frac{B}{B^2 - \omega_0^2}
\,.
\end{equation}
This correction (the Lamb shift) has different signs for fast ($\omega_0 > B$) 
and slow ($\omega_0 < B$) oscillators. As one can see from 
Eqs.~(\ref{Eq:dEup}), 
(\ref{Eq:dEdn}), this result can be understood in terms of the level 
repulsion~\cite{FrankW-Osc}: the perturbation couples the level $\ket{\up,0}$ 
to 
$\ket{\dn,1}$ and $\ket{\dn,0}$ to $\ket{\up,1}$. The levels of the latter 
pair are closer, and the coupling has a stronger effect on their energies. 
They repel 
each other due to the coupling, thus reducing the distance between 
$\ket{\up,0}$ 
and $\ket{\dn,0}$ for $\omega_0 > B$ and increasing it for $\omega_0 < B$.

%%%%%%%%%%%%%%%%%%%%%%%%%%%%%%%%%%%%%%%%%%%%%%%%%%%
\subsection{Second-order perturbative analysis}
\label{sect:goldenrule}

In this section we find the Lamb shift using the lowest-, second-order 
perturbative analysis. In the Hamiltonian (\ref{Eq:Ham}) we treat the coupling 
term ${\cal V}= -\half X \sigma_z$ as a perturbation: ${\cal H}= {\cal 
H}_0 + {\cal V}$. The eigenstates of ${\cal H}_0$ are $\ket{\alpha,i}$, 
where 
$\alpha=\up_B$/$\dn_B$ denotes the eigenstates of the spin without 
dissipation, with 
the spin direction parallel or antiparallel to the filed $\bi{B}$, and $i$ 
denotes eigenstates of the environment. The perturbation theory gives for the 
corrections to their eigenenergies:
\begin{eqnarray}
E_{\alpha,i}^{(2)} = -
\sum_{\beta,j} \frac{\stbrkt{ \bra{\alpha,i}{\cal 
V}\ket{\beta,j}}^2}{E^{(0)}_{\beta} + 
E_j^{(0)}-E^{(0)}_{\alpha} - E^{(0)}_i-\rmi 0} 
\,.
\label{Eq:d2Eni}
\end{eqnarray}
For ${\cal V} = -\half X \sigma_z$ we notice that 
$\bra{\up_B} \sigma_z \ket{\up_B}^2 =
\bra{\dn_B} \sigma_z \ket{\dn_B}^2 =\cos^2\theta$ and
$\bra{\up_B} \sigma_z \ket{\dn_B}^2 =
\bra{\dn_B} \sigma_z \ket{\up_B}^2 =\sin^2\theta$, and find
for the environment-averaged quantities $E_{\alpha}^{(2)} \equiv \sum_i \rho_i\, 
E_{\alpha,i}^{(2)}$ (see the discussion of these quantities at the end of this 
subsection):
\begin{equation}
E_{\up}^{(2)}  =
-\frac{\cos^2\theta}{4}
\sum_{i,j}
\frac{\rho_i\,|\bra{i}X\ket{j}|^2}{E_j^{(0)}-E_i^{(0)}-\rmi 0}
-\frac{\sin^2\theta}{4}
\sum_{i,j}
\frac{\rho_i\,|\bra{i}X\ket{j}|^2}{B+E_j^{(0)}-E_i^{(0)}-\rmi 0}
\,.
\label{Eq:E_up}
\end{equation}
The correction to $E_{\dn}$ is obtained by substituting $B \rightarrow 
-B$ into the above equation.
Now using the identity
\begin{equation}
\frac{1}{E-\rmi 0}=\rmi\,\int_{0}^{\infty} dt\,e^{-i(E-\rmi 0)t}\,,
\end{equation}
we rewrite Eq.~(\ref{Eq:E_up}) as
\begin{equation}
E_{\up}^{(2)}  = -\frac{\rmi}{4} \int_{0}^{\infty}
dt\, \langle X(t)X(0)\rangle
\left( \cos^2\theta + \sin^2\theta e^{-iBt} \right) e^{-0 t}
\,,
\label{Eq:E_up_time}
\end{equation}
where we have used the relation
\begin{equation}
\langle X(t)X(0)\rangle = \sum_{i,j} \rho_i 
\bra{i}X\ket{j}\bra{j}X\ket{i}e^{-i(E_j - E_i)t}
\,.
\end{equation}
In terms of the the Fourier transform
$\langle X^2_\nu \rangle \equiv \int dt\, \langle X(t)X(0)\rangle\,e^{i\nu t}$
we obtain
\begin{equation}
\label{Eq:E_up_nu}
E_{\up}^{(2)} = -\frac{1}{4} \cos^2\theta\,
\int \frac{d\nu}{2\pi}\, \frac{\langle X^2_{\nu} \rangle}{\nu-\rmi 0}
-\frac{1}{4} \sin^2\theta\,
\int \frac{d\nu}{2\pi}\, \frac{\langle X^2_{\nu} \rangle}{\nu+B-\rmi 0}
\,.
\end{equation}
For the Lamb shift $E_{\rm Lamb}^{(2)} \equiv \Ree (E_\dn^{(2)} - E_\up^{(2)})$ 
this gives a principal value integral
\begin{equation}
E_{\rm Lamb}^{(2)} =
\half\sin^2\theta\, \pvint \frac{d\nu}{2\pi}
\frac{S_X(\nu)}{B-\nu}
=
B \sin^2\theta\, \pvint_0^\infty \frac{d\nu}{2\pi}\,
\frac{S_X(\nu)}{B^2-\nu^2} \,,
\label{Eq:Lamb_Shift}
\end{equation}
where
\begin{equation}
S_X(\nu) \equiv \half(\langle X^2_{\nu}\rangle + \langle
X^2_{-\nu}\rangle) =
\half\int dt \, \langle [X(t),X(0)]_+\rangle\,e^{i\nu t} \,.
\end{equation}
Thus the Lamb shift is expressed in terms of the symmetrized correlator $S_X$ 
and is insensitive to the antisymmetric part of the noise spectrum.

As one can see from Eq.~(\ref{Eq:Lamb_Shift}), in agreement with the discussion 
in the previous section, the high-frequency
noise ($\nu > B$) reduces the energy gap between
the spin states~\cite{Leggett87}, while the low frequency
modes ($\nu < B$) increase the energy gap.

Similarly, from Eq.~(\ref{Eq:E_up_nu}) one can evaluate the dephasing time:
\begin{equation}
\frac{1}{T_2} = - \Imm (E_\up^{(2)} + E_\dn^{(2)}) =
\frac{\cos^2 \theta}{4} 
S_X(\nu=0) + \frac{\sin^2 \theta}{4} S_X(\nu=B)
\,.
\label{Eq:T2}
\end{equation}
This expression correctly reproduces the contribution of the 
transverse fluctuations ($\propto\sin^2\theta$) to the dephasing rate, but 
underestimates the longitudinal contribution  ($\propto\cos^2\theta$) by a 
factor of two (cf.~Ref.~\cite{Bloch_Derivation,Redfield_Derivation,WeissBook}). 
One can show that an accurate 
evaluation of this contribution, as well as the analysis of the relaxation, 
requires taking into account corrections to the eigenstates, and not only to the 
eigenenergies (\ref{Eq:d2Eni}). More precisely, our calculation of the 
corrections to the eigenenergies in this subsection corresponds to evaluation 
only of the four left diagrams in Fig.~7 of Ref.~\cite{Our_Erice}; the term 
$\rmi 0$ in the denominators allows one to find also the outgoing transition 
rates {\it from} the eigenstates (and the respective contribution, $\propto 
\sin^2\theta$, to dephasing) but only the part of the `pure-dephasing' rate, 
$\propto \quarter \cos^2\theta$. Analysis of the two remaining diagrams in 
Fig.~7 and those in Fig.~6 allows one to find also the pure dephasing rate (as 
well as the incoming transition rates, the latter though do not require an extra 
evaluation due to probability conservation).

%%%%%%%%%%%%%%%%%%%%%%%%%%%%%%%%%%%%%%%%%%%%%%%%%%%
\subsection{From Lamb shift to Berry phase}
\label{sec:LStoBP}

So far we have analyzed the environment-induced correction to the level
splitting (the Lamb shift).  Using the results above one can evaluate also the
environment-induced correction to the Berry phase for a slow cyclic variation of
the magnetic field $\bi{B}$
\cite{wg-moriond,wg-prl,wmsg-geometricBP,wg-longpaper}.

\begin{figure}
\centerline{\hbox{\psfig{figure=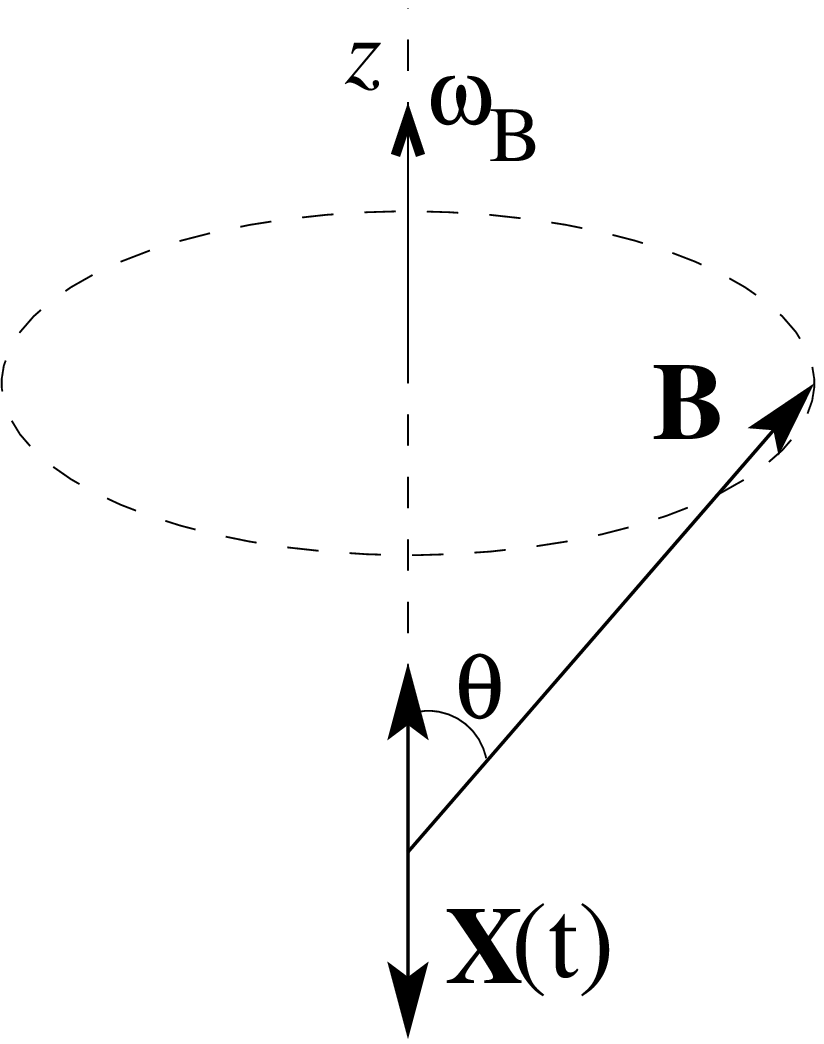,width=0.35\columnwidth}}}
\caption{\label{Fig:BerryExp}}
\end{figure}

Indeed, consider the simplest case of conic variations of the field around the
$z$-axis (to which the environment is coupled), as shown in 
Fig.~\ref{Fig:BerryExp}: the field varies at a constant rate, with the low 
angular velocity $\omega_{\rm B}$, and traverses the circle after the period 
$t_P\equiv 2\pi/\omega_{\rm B}$. The analysis of the spin dynamics is
considerably simplified by going to the frame, 
rotating with the angular
velocity $\omega_{\rm B}\hat z$, where $\hat z$ is the unit vector along 
the $z$-axis. 
In this frame the spin is subject to the 
fluctuating field $X\hat z$ and the field $\bi{B}+\omega_{\rm B}\hat z$, which 
is {\it stationary}. Thus, in this frame one can use the results of the 
analysis
above to obtain the Lamb shift, if one substitutes $\bi{B}$ by 
$\bi{B}+\omega_{\rm B}\hat z$. In other words, the correction to the Lamb shift
associated with the variation of the field $\bi{B}$ in time, is given by taking
the derivative $\omega_{\rm B}\partial_{B_z}$ of the Lamb shift
(\ref{Eq:Lamb_Shift}) and multiplying by the period of variation, $t_P$. 
After a full period the basis of the rotating frame makes a complete circle 
and returns to
its initial position, i.e. coincides with the laboratory frame's basis. Hence
the phases accumulated in the rotating and laboratory frames coincide, and it 
is
sufficient to evaluate it in the rotating frame. Thus, one finds the
environment-induced correction to the Berry phase to be
\begin{equation} \delta
\Phi_{BP} = 2\pi\, \frac{\partial E_{\rm Lamb}(\bi{B})}{\partial B_z}\,.
\label{Eq:BPfromLamb}
\end{equation}

%%%%%%%%%%%%%%%%%%%%%%%%%%%%%%%%%%%%%%%%%%%%%%%%%%%%
%%%%%%%%%%%%%%%%%%%%%%%%%%%%%%%%%%%%%%%%%%%%%%%%%%%%

Taking the derivative of Eq.~(\ref{Eq:Lamb_Shift}), we find:
\begin{equation}
\delta^{(2)}\Phi_{BP} =
\cos\theta \sin^2\theta\,
\pvint d\nu\, \frac{S_X(\nu)(2\nu-3B)}{2B(B-\nu)^2}
\,.
\label{Eq:dBP}
\end{equation}
(Notice the convention: this expression gives the correction to the relative 
Berry phase between the spin-up and spin-down states, rather than to the 
phases 
of each of these states.) As for the Lamb shift, the contributions of the high-
and low-frequency fluctuations are of opposite signs. For the Berry phase the
contribution changes sign at $\nu=3B/2$.

In passing we note that this analysis can be generalized to
an arbitrary (but adiabatic) path $\bi{B}(t)$,
this enables one to see that the correction to the Berry phase is
geometric, but that its geometric nature is very different from
the Berry phase of an isolated spin-half \cite{wmsg-geometricBP}.

In Section~\ref{sec:classical} we shall find exactly the same
expression for the Lamb shift and therefore for the Berry
phase in the case of classical environment.

%%%%%%%%%%%%%%%%%%%%%%%%%%%%%%%%%%%%%%%%%%%%%%%%%%%
\subsection{High-frequency noise: renormalization of the transverse 
$\bi{B}$-field}
\label{sec:hf-renorm}

Consider now the influence of the high-frequency fluctuations in the 
environment 
only ($\nu\gg B$). Since the frequencies of the fluctuations are much higher 
than the typical spin-dynamics frequencies, one may eliminate these 
high-frequency fluctuations using the adiabatic (Born-Oppenheimer) 
approximation, as described, e.g., by Leggett et al. \cite{Leggett87}.

Indeed, consider the spin-boson model, with the Hamiltonian
\begin{equation}
{\cal H} = -\half(\bi{B}+X\hat z)\bi{\sigma} + {\cal H}_{\rm env}\,,
\label{Eq:HXhigh}
\end{equation}
where $X=\sum_i c^{\phantom{\dagger}}_i (a^\dagger_i+a^{\phantom{\dagger}}_i)$
and ${\cal H}_{\rm env}= \sum_i \omega^{\phantom{\dagger}}_i a^\dagger_i
a^{\phantom{\dagger}}_i$. Let us ignore the low-frequency oscillators and
focus on those at high frequencies $\nu\gg B$. These fast oscillators adjust
almost instantaneously to the slowly varying spin state. For the last
two terms of the Hamiltonian (\ref{Eq:HXhigh}) two lowest-energy states are
$\ket{\tilde\up}=\ket{\up}\prod_i\ket{g^\up_i}$ and
$\ket{\tilde\dn}=\ket{\dn}\prod_i\ket{g^\dn_i}$. Here $\ket{g^\up_i}$ denotes
the ground state of the $i$th oscillator corresponding to the spin state
$\ket{\up}$, i.e. the ground state of $\omega^{\phantom{\dagger}}_i a^\dagger_i
a^{\phantom{\dagger}}_i + c^{\phantom{\dagger}}_i
(a^\dagger_i+a^{\phantom{\dagger}}_i)$, and $\ket{g_i^\dn}$ is defined
similarly; further eigenstates of the last two terms are separated by a gap
$\sim\nu$.

Consider now the matrix elements of the
first term $-\half\bi{B}\bi{\sigma}$ in this two-state 
low-energy subspace; one finds
that its transverse component is suppressed by the factor
\begin{equation}
\prod_i \braket{g_i^\up}{g_i^\dn} = \prod_i
\exp(- c_i^2 /2 \omega_i^2) = \exp\left( -\int_0^\infty \frac{d\nu}{2\pi}
\frac{J_X(\nu)}{\nu^2} \right)\,,
\label{Eq:BxSupT=0}
\end{equation}
where $J_X(\nu)\equiv \pi\sum_i c_i^2 \delta(\nu-\omega_i)$ is the spectral
density of the oscillator bath. At a finite temperature $T$ each high-frequency
oscillator remains in its thermal equilibrium state (subject to the spin
state), rather than the ground state, and on the rhs of Eq.~(\ref{Eq:BxSupT=0})
the spectral density $J_X(\nu)$ is
replaced by the thermal noise power $S_X(\nu) = J_X(\nu) \coth(\nu/2k_{\rm
B}T)$.

Thus the role of the high-frequency oscillators is to suppress the transverse
field component (in other words, the transverse $g$-factor). If we are
interested only in the contribution to the level spacing (the Lamb shift), 
one should consider
only the longitudinal ($\pll\bi{B}$) part of the renormalization, i.e. multiply
the result by $\sin\theta$, to obtain Eq.~(\ref{Eq:Lamb_Shift}).

%%%%%%%%%%%%%%%%%%%%%%%%%%%%%%%%%%%%%%%%%%%%%%%%%%%
\subsection{Effective-action analysis}
\label{sec:Effective_Action}

One can study the spin dynamics integrating
out the environment and using the effective action  for the spin. 
We derive the effective action using the Feynman-Vernon-Keldysh technique.
For the interaction $-Xs_z$ with the $z$-component of the spin,
the effective action (the influence functional) reads
\begin{eqnarray}
\rmi \Phi_{\rm infl}  = -\frac{1}{2} \int_{C_K} dt\int_{C_K} dt'\,
s_z(t)\cdot s_z(t') \, [\,\rmi G_X(t,t')\,]
\,,
\label{Eq:Seff}
\end{eqnarray}
where we assumed the Gaussian statistics 
of $X$, and defined the Green function $G_X$
as $\rmi G_X(t,t') = \langle T_{C_K} X(t)X(t')\rangle$. The time ordering here 
refers to the Keldysh time contour $C_K$, and in Eq.~(\ref{Eq:Seff}) we 
integrate over $C_K$; accordingly each of the time dependent variables 
assumes a `Keldysh
index' $u,d$ indicating the upper/lower branch of this contour.

After the Keldysh rotation one obtains the influence functional in terms of the 
classical and quantum components, $s_z^c \equiv (s_z^u + s_z^d)/2$
and $s_z^q \equiv s_z^u - s_z^d$:
\begin{equation}
\Phi_{\rm infl}  = - \int dt dt'
\left[
s_z^q(t) G_X^{\rm R}(t-t') s_z^c(t') + 
\frac{1}{4}\, s_z^q(t) G_X^{\rm K}(t-t') s_z^q(t')
\right]
\,,
\label{Eq:Seff_rotated}
\end{equation}
in terms of the retarded and Keldysh Green functions, $G_X^{\rm R} \equiv -\rmi
\theta(t-t') \langle [X(t),X(t')]_- \rangle$ and $G_X^{\rm K} \equiv -\rmi
\langle [X(t),X(t')]_+ \rangle = -2\rmi S_X(t-t')$.

For classical noise $X$ the commutator in the definition of $G^{\rm R}$ 
vanishes, and one finds
\begin{equation}
\Phi_{\rm infl}^{\rm class}  = \frac{\rmi}{2} \, \int dt\int dt'\,
\,s_z^q(t) \,S_X(t-t')\, s_z^q(t')
\,.
\label{Eq:Seff_classical}
\end{equation}

The results~(\ref{Eq:Lamb_Shift}), (\ref{Eq:dBP}) for the Lamb shift and the
Berry phase involve only $S_X$ and not the antisymmetrized correlator. Hence 
for
the analysis of these quantities it should be sufficient to use the functional
(\ref{Eq:Seff_classical}). Alternatively, one may consider a problem with a
classical random field $\bi{X}(t)$ to reproduce these results. In the next
section we perform the corresponding analysis.

%%%%%%%%%%%%%%%%%%%%%%%%%%%%%%%%%%%%%
\section{The Classical  Model}
\label{sec:classical}

In this section we analyze the dynamics of a spin subject to a classical random
field and derive the equation of motion for the spin dynamics (the
spin-evolution operator), averaged over the fluctuations. Following the
discussion of the case with quantum fluctuations, we first analyze the dynamics
in a stationary field $\bi{B}$ and a random field; exactly as in the quantum
case one can reduce the analysis of the dissipative corrections to the Berry
phase accumulated over a conic loop to the problem with a stationary field by
going over to a rotating frame.

As we have demonstrated above, 
in the quantum problem the results for the corrections 
to the phase and dephasing, associated with the controlled dynamics of the
magnetic field, involve only the symmetric part of the noise correlator, one 
expects that the results for these quantities in the classical problem, 
expressed 
in terms of the noise power, would coincide with the quantum results. Indeed, 
we 
find this relation below.

Specifically, we analyze the following problem: a spin $\bi{S}$ is coupled to a 
controlled magnetic field $\bi{B}$ (stationary for now, but to be varied slowly 
in a Berry-phase experiment) and a randomly fluctuating field 
$\bi{X}(t)$, which we treat as a random variable
with the correlation function given by $S_X(t)$.
Its 
dynamics is governed by the Larmor equation:
\begin{equation}
\dot{\bi{S}}=[\bi{B} + \bi{X} (t)] \times \bi{S} \,.
\end{equation}
This equation can be used to describe the dynamics of either a
classical spin or the average spin value (i.e. the density matrix) of a
spin-$1/2$.

As we discussed in the Introduction, we assume that the noise is weak and
short-correlated, i.e., that considerable dissipative contributions to the spin
dynamics arise on time scales much longer than the typical correlation time
$\tau_{\rm c}$ of the noise.  Below we discuss the influence of the low- and
high-frequency fluctuations on the (classical) spin dynamics and recover the
results of the quantum analysis above.  Further, using the result for the
low-frequency contribution we obtain the correction to the Berry phase from the
environmental fluctuations at all frequencies.

%%%%%%%%%%%%%%%%%%%%%%%%%%%%%%%%%%%%%%%%%%%%%%%%%%%%%%%%%%5
\subsection{Low-frequency noise: Lamb shift}

Consider first the effect of a slowly fluctuating random field $\bi{X} = X\hat 
z$. Similar to the quantum-mechanical analysis in Section~\ref{sec:quantum} we 
begin with the case of harmonic fluctuations (of random amplitude) and purely 
transverse noise ($\bi{B}=B\hat x$, i.e. $\theta=\pi/2$). Consider fluctuations 
$X=c_\nu\cos(\nu t)$ at a low frequency $\nu\ll B$, during a time interval 
$\delta t$. To evaluate the evolution operator, we analyze the
dynamics in a reference frame $(\hat\xi,\hat\eta,\hat\zeta)$ fluctuating
together with the field (with the $\zeta$-axis along $\bi{B}+\bi{X}(t)$ and
the $\eta$-axis, for instance, orthogonal to $\bi{B}$ and $\bi{X}$). Since the
fluctuating angular velocity of this frame's rotation is negligible,
$\sim c_\nu \nu/B \ll c_\nu$, the effective magnetic field in
this frame $|\bi{B}+\bi{X}(t)|\hat \zeta$ points along the $\zeta$-axis. Thus
the dynamics reduces to rotation about this axis by the angle $\phi(t) =
\int_t^{t+\delta t} d\tau |\bi{B}+\bi{X}(\tau)| \approx \int_t^{t+\delta t}
d\tau (B+ X^2(\tau)/2B)$, where $B=|\bi{B}|$. Averaging the transverse spin 
component
$S_x+iS_y \propto e^{i\phi(t)}$ one finds a lowest-order
contribution to the phase factor, $\delta t \langle X^2 \rangle /2
B$, i.e. a Lamb shift $c_\nu^2/4 B$ (we assumed $\delta t$ much longer than the 
period of oscillations, $1/\nu$).

In principle, the evolution in the laboratory frame differs from that in the 
rotation frame. Transformation to/from the rotation frame at the beginning and 
the end of the time interval introduces corrections to the evolution operator 
or 
order $c_\nu/B$. This is however a negligible boundary contribution. Indeed, 
for 
a sufficiently long time interval $\delta t\gg 1/c_\nu$ the phase shift due to 
the Lamb shift, of order $c_\nu^2\delta t/B$, is much larger (but still small, 
as long as $\delta t\ll B/c_\nu^2$).

Similar results hold for more general low-frequency fluctuations, non-harmonic 
and with arbitrary direction $\theta$. Indeed, in the same rotating frame
the dynamics reduces to rotation about the $\zeta$-axis by the angle $\phi(t) =
\int_t^{t+\delta t} d\tau |\bi{B}+\bi{X}(\tau)| \approx \int_t^{t+\delta t}
d\tau (B+ X_\parallel(\tau)+ X_\perp^2(\tau)/2B)$, where
$X_\parallel=X\cos\theta$, $X_\perp=X\sin\theta$ are the longitudinal and
transverse components of $\bi{X}$ (relative to $\bi{B}$). Averaging the
transverse spin component $\propto e^{i\phi(t)}$ one finds, apart from
dephasing, a lowest-order contribution to the phase factor, $\delta t \langle
\bi{X}_\perp^2 \rangle /2 B$, and hence the Lamb shift
\begin{equation}
\delta E = \sin^2\theta\int \frac{d\nu}{4\pi} \frac{S_X(\nu)}{B}\,,
\label{Eq:LambShiftLowFreq}
\end{equation}
where $\theta$ is the angle between $\bi{B}$ and the direction $\hat z$ of the
noise. This result coincides with the low-frequency contribution in
Eq.~(\ref{Eq:Lamb_Shift}).

%%%%%%%%%%%%%%%%%%%%%%%%%%%%%%%%%%%%%%%%%%%%%%%%%%%%%%%%%%%%%%%%
\subsection{From low frequencies to all frequencies}

The expression (\ref{Eq:LambShiftLowFreq}) and the symmetry of the problem
suggests a way to find the contribution of all, not only slow, modes in the
environment to the Lamb shift (and later to the Berry phase). Indeed, we 
discuss
weak short-correlated noise, i.e. such that its contribution to the dynamics on
time scales or order $\tau_{\rm c}$ is small. Contributions from different time
intervals $\sim\tau_{\rm c}$ are uncorrelated and add up independently. Hence
in the evaluation of the (real and imaginary) contribution of such a short
interval to the evolution frequencies (the Lamb shift, the dephasing and
relaxation rates) it is enough to consider the lowest nonvanishing, i.e. second
order.

The symmetry of the problem can be used to analyze the structure of such a
second-order contribution. The spin-rotational symmetry (about the
$\bi{B}$-field's direction) and the time-translational symmetry imply that (i)
the longitudinal and transverse fluctuations, $X_\perp$ and $X_\pll$, do not
interfere and may be considered separately; (ii) it is convenient to expand the 
transverse fluctuating field in circularly polarized harmonic modes, and the 
latter contribute independently.

The longitudinal noise gives rise to the pure dephasing (and only the low 
frequencies $\lesssim 1/T_2$ contribute), without affecting the level 
splitting.
As for the transverse noise, for a single circularly polarized mode at 
frequency 
$\nu$ it is convenient to analyze its contribution in the spin frame, rotating 
at frequency $\nu$ around the field $\bi{B}$. In this frame the Larmor field is
$B-\nu$ in the direction of $\bi{B}$, and the fluctuating circularly polarized 
mode is slow. 
Applying to this mode Eq.~(\ref{Eq:LambShiftLowFreq}), going back to the 
laboratory frame and adding up contributions of all modes we arrive at the 
expression for the correction to the Larmor frequency:
\begin{equation}
\delta E = \sin^2\theta \pvint \frac{d\nu}{4\pi} \frac{S(\nu)}{B-\nu}\,.
\label{Eq:LambShiftAllFreq}
\end{equation}
It is thus this result which needs to be compared with the quantum correction
(Lamb shift) of the previous section.
Symmetrization of the integral in Eq.~(\ref{Eq:LambShiftAllFreq}) w.r.t. to 
$\nu$ brings it to the form of Eq.~(\ref{Eq:Lamb_Shift}). Notice that
regularization of this expression via the introduction of $+i0$ in the 
denominator allows us also to recover the imaginary part of the Lamb shift, 
i.e. 
a contribution to the dephasing rate.

%%%%%%%%%%%%%%%%%%%%%%%%%%%%%%%%%%%%%%%%%%%%%%%%%%%%%%%%
\subsection{High frequencies}

%%%%%%%%%%%%%%%%%%%%%%%%%%%%%%%%%%%%
\begin{figure}
\centerline{\hbox{\psfig{figure=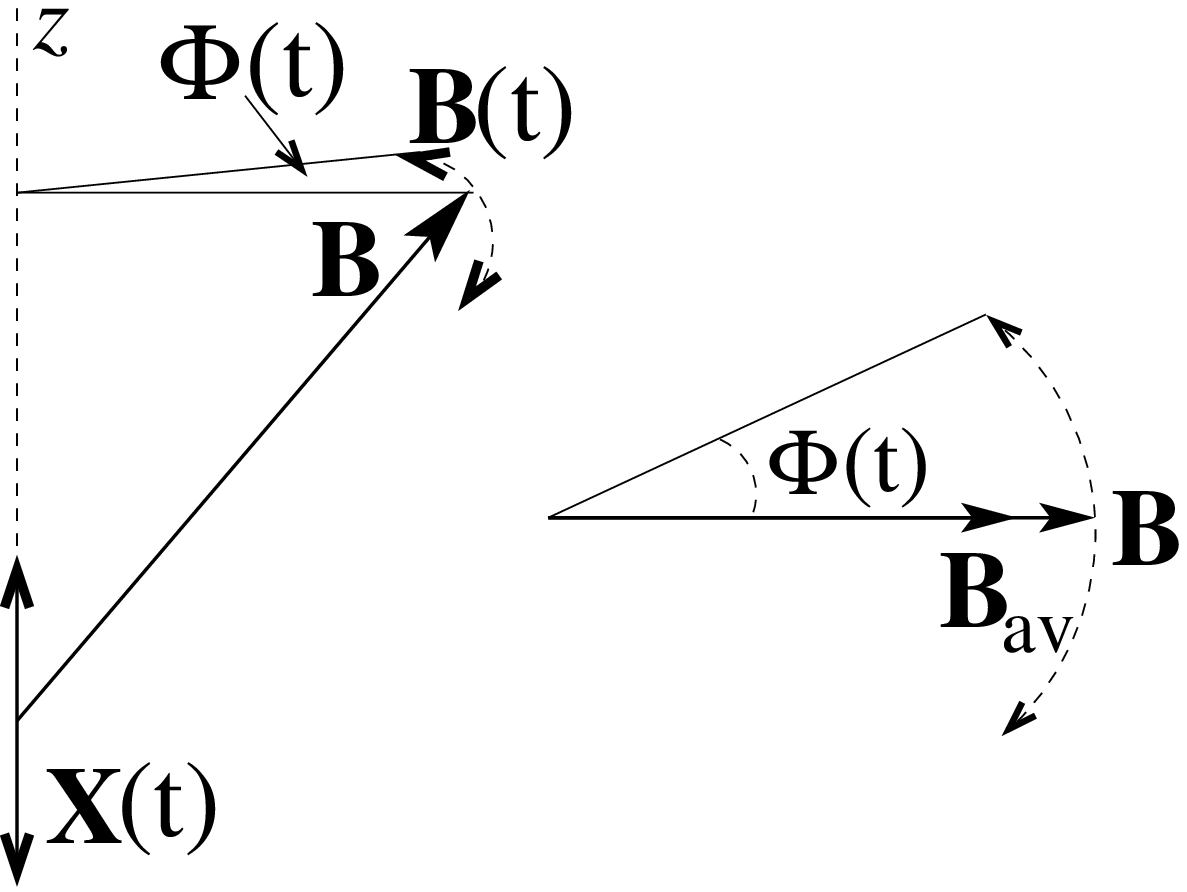,width=0.35\columnwidth}}}
\caption{\label{Fig:HighFreqClass}}
\end{figure}
%%%%%%%%%%%%%%%%%%%%%%%%%%%%%%%%%%%%

Although Eq.~(\ref{Eq:LambShiftAllFreq}) describes the contribution of all 
frequencies, it is interesting to discuss specifically the limit of high 
frequencies. In this subsection we provide an argument which parallels the 
result of subsection~\ref{sec:hf-renorm}: the high-frequency fluctuations 
$(\nu\gg B$) suppress the transverse ($\perp\hat z$) component of the 
$\bi{B}$-field.

Indeed, to solve for the dynamics in the presence of high-frequency 
fluctuations 
in a fixed direction, $X(t)\hat z$, and the static field $\bi{B}$, let us 
analyze the dynamics in the frame that rotates about the $\hat z$-axis with  
angular velocity $X(t)$, i.e. differs from the lab frame by a rotation by the 
fluctuating angle $\Phi(t)=\int_0^t X(\tau)d\tau$. The rotation of this frame 
is
chosen to exactly compensate for the field $X(t)\hat z$, and the Larmor field 
$\bi{B}(t)$ in this frame is just the $\bi{B}$-field, but now fluctuating due 
to 
the frame's rotation as shown in Fig.~\ref{Fig:HighFreqClass}. The spin 
dynamics 
is governed by the Larmor equation $\dot{\bi{S}}=\bi{B}(t)\times \bi{S}$, and 
the value of the spin changes considerably only on time scales of order $1/B$, 
during which many fluctuations occur. Looking at the dynamics on
intermediate time scales, between $1/\nu$ and $1/B$, one finds that the spin
dynamics is governed by the value of the $\bi{B}$-field averaged over fast 
fluctuations. The averaging affects only the horizontal (orthogonal to $z$)
component of the $\bi{B}$-field. The direct evaluation shows that the 
horizontal 
component is suppressed exactly by the factor 
$\exp[-\int\limits_{0}^{\infty}(d\nu/2\pi) 
S_X(\nu)/\nu^2]$ (cf.~Eq.~(\ref{Eq:BxSupT=0})).
For instance, for a single mode at frequency $\nu$ we have $X(t)=2 
X_\nu\cos(\nu t)$ and $\Phi(t)=2 X_\nu\sin(\nu t)/\nu$; then the transverse 
component of the field is suppressed by the factor $1-\langle \Phi^2\rangle/2$, 
and $\langle\Phi^2\rangle/2=\langle X_\nu^2\rangle/\nu^2$.
This evaluation of the dynamics in the rotating  frame relies on the small 
parameter $B/\nu$.

The spin-evolution operator (before averaging) $\hat O_{\rm lab}(t,t')$ in the 
laboratory frame is related to that in the rotating frame, $\hat O_{\rm
lab}(t,t') = \hat O_z(-\Phi(t)) \hat O_{\rm rot}(t,t') \hat O_z(\Phi(t'))$, via  
the transformation $O_z(\Phi(t))$ from the lab frame to the rotating frame. 
However, this transformation $\hat O_z(\Phi(t))$ at the beginning and at the 
end 
of the evolution is close to the identity operator, and taking it into account 
adds only a boundary effect, which does not grow with the size of the time 
interval and is therefore negligible.

%%%%%%%%%%%%%%%%%%%%%%%%%%%%%%%%%%%%%%%%%%%%%%%%%%%%%%%%
\subsection{Berry phase under classical noise}
\label{sec:classBerry}

To find a dissipation-induced correction to the Berry phase we may use the same 
approach as in Section~\ref{sec:LStoBP}: first, we find
the Lamb shift for a stationary field $\bi{B}$ and then evaluate the Berry 
phase
using the relation (\ref{Eq:BPfromLamb}). In this way we find the same 
expression (\ref{Eq:dBP}) for the Berry phase.

%%%%%%%%%%%%%%%%%%%%%%%%%%%%%%%
\section{Conclusions}
\label{sec:concl}

In this paper we have derived expressions for the environment-induced 
correction 
to the Berry phase, for a spin coupled to an environment. On one hand, we
presented a simple quantum-mechanical derivation for the case when the 
environment is treated as a separate quantum system. On the other hand, we 
analyzed the case of a spin subject to a random classical field. The 
quantum-mechanical derivation provides a result which is insensitive to the 
antisymmetric part of the random-field correlations.  In other words, the 
results for the Lamb shift and the Berry phase are insensitive to whether the
different-time values of the random-field operator commute with each other or 
not. This observation gives rise to the expectation that for a random classical 
field, with the same noise power, one should obtain the same result. 
For the quantities at hand, our analysis outlined above involving 
classical randomly fluctuating fields has confirmed this expectation.

Furthermore, we provided simple arguments, which allow one to understand the 
contribution of fluctuations in various frequency ranges (below and above the 
Larmor frequency).

We thank M.~Cholascinski, R.~Chhajlany, G.~Sch\"on for discussions.
This work was supported by the German-Israeli foundation (GIF), by the 
Minerva foundation, by the Israel Science Foundation (ISF) of the Israel 
Academy, by the Center for Functional Nanostructures (CFN) of the DFG, and by 
the Landesstiftung BW.
R.W. was also supported by EPSRC fellowship GR/R44836/01 
during the latter part of this work. 
Y.M. was supported by the S.~Kovalevskaya award of the Humboldt foundation,
the BMBF, and the ZIP programme of the German government.

%%%%%%%%%%%%%%%%%%%%%%%%%%%%%%%%%%%%%%%%%%%%%%%%

\bibliographystyle{apalike}
\chapbblname{gefen}
\chapbibliography{gefen}

\end{document}